\documentclass[%
 aip,
 amsmath,amssymb,
 reprint,%
]{revtex4-1}

\usepackage{graphicx}
\usepackage{dcolumn}
\usepackage{bm}
\usepackage[mathlines]{lineno}

\usepackage[utf8]{inputenc}
\usepackage[T1]{fontenc}
\usepackage{mathptmx}
\usepackage{etoolbox}
\usepackage{xcolor}

\newcommand{\mypd}[2]{\frac{ \partial #1}{ \partial #2}}

\newcommand{\mygvec}{\mathbf{g}}

\makeatletter
\def\@email#1#2{%
 \endgroup
 \patchcmd{\titleblock@produce}
  {\frontmatter@RRAPformat}
  {\frontmatter@RRAPformat{\produce@RRAP{*#1\href{mailto:#2}{#2}}}\frontmatter@RRAPformat}
  {}{}
}%
\makeatother
\begin{document}

\preprint{AIP/123-QED}

\title[Electron-Capture Elements in Thermonuclear Supernovae]{The Production of Electron-Capture Elements in Thermonuclear Supernovae: Theory vs. Observations}

\newcommand{\FSU}{\affiliation{Department of Physics, Florida State
    University, 77 Chieftan Way, Tallahassee, FL 32306, USA}}
\newcommand{\PSI}{\affiliation{Planetary Science Institute, 1700 East Fort
  Lowell Road, Suite 106,Tucson, AZ 85719-2395 USA}}
\newcommand{\HS}{\affiliation{Hamburger Sternwarte, Gojenbergsweg 112, 21029 Hamburg, Germany}}
\newcommand{\IFA}{\affiliation{Institute for Astronomy, University of Hawai’i at Manoa, 2680 Woodlawn Dr., Hawai’i, HI 96822, USA}}
\newcommand{\VT}{\affiliation{Department of Physics, Virginia Tech,
    850 West Campus  Drive, Blacksburg VA, 24061, USA}}
\newcommand{\GRFP}{\altaffiliation{National Science Foundation Graduate Research Fellow}}
\newcommand{\FINESST}{\altaffiliation{NASA FINESST Future Investigator}}
\newcommand{\NHFPE}{\altaffiliation{NHFP Einstein Fellow}}
\newcommand{\UIUC}{\affiliation{Department of Astronomy, University of Illinois Urbana-Champaign, 1002 West Green Street, Urbana, IL 61801, USA}}
\newcommand{\NSFSIMS}{\affiliation{NSF-Simons AI Institute for the Sky (SkAI), 172 E. Chestnut St., Chicago, IL 60611, USA}}

\newcommand{\STSci}{\affiliation{Space Telescope Science Institute, 3700 San Martin Drive, Baltimore, MD 21218-2410, USA}}
\newcommand{\Carnegie}{\affiliation{Observatories of the Carnegie
    Institution for Science, 813 Santa Barbara St., Pasadena, CA 91101, USA}}
\newcommand{\MSU}{\affiliation{Department of Physics \& Astronomy,
    Michigan State University, East Lansing, MI, USA}}
\newcommand{\TAMU}{\affiliation{George P. and Cynthia Woods Mitchell
    Institute for Fundamental Physics and Astronomy,
    Department of Physics and Astronomy, Texas 
             A\&M University, College Station, TX 77843, USA}}
\newcommand{\IALP}{\affiliation{Instituto de Astrof\'isica de La Plata
    (IALP), CONICET, Paseo del Bosque S/N, B1900FWA La Plata, Argentina}}
\newcommand{\LaPlata}{\affiliation{Facultad de Ciencias Astron\'omicas
    y Geof\'isicas Universidad Nacional de La Plata, Paseo del Bosque,
    B1900FWA, La Plata, Argentina}}
\newcommand{\WPI}{\affiliation{Kavli Institute for the Physics and
    Mathematics of the Universe (WPI), The University of Tokyo,
    Kashiwa, 277-8583 Chiba, Japan}} 

\newcommand{\ICE}{\affiliation{Institute of Space Sciences (ICE,
    CSIC), Campus UAB, Carrer de Can Magrans, s/n, E-08193 Barcelona, Spain}}

\newcommand{\IEEC}{\affiliation{Institut d’Estudis Espacials de
    Catalunya (IEEC), E-08034  Barcelona, Spain}} 

\newcommand{\LCO}{\affiliation{Las Campanas Observatory, Carnegie
    Observatories, Casilla 601, La Serena, Chile}} 

\newcommand{\Aarhus}{\affiliation{Department of Physics and Astronomy,
    Aarhus University, Ny  Munkegade 120, DK-8000 Aarhus C, Denmark.}} 

\newcommand{\OU}{\affiliation{Homer L.~Dodge Department of Physics and
  Astronomy, University of Oklahoma, 440 W. Brooks, Rm 100, Norman, OK
  73019-2061}}  

\newcommand{\UCSC}{\affiliation{Department of Astronomy and Astrophysics,
  University of California, Santa Cruz, CA 95064, USA}} 
\newcommand{\Melbourne}{\affiliation{School of Physics, The University of
  Melbourne, VIC 3010, Australia}}

\newcommand{\LPNHE}{\affiliation{LPNHE, (CNRS/IN2P3, Sorbonne
  Universit\'e, Universit\'e Paris Cit\'e), Laboratoire de Physique
  Nucl\'eaire et de Hautes \'Energies, 75005, Paris, France}}

\newcommand{\Princeton}{\affiliation{Princeton University, 4 Ivy Lane,
    Princeton, NJ 08544, USA}}

\newcommand{\Berkeley}{\affiliation{Department of Astronomy,
    University of California, Berkeley, CA 94720-3411, USA}}

\newcommand{\Tsinghua}{\affiliation{Physics Department, Tsinghua
    University, Beijing, 100084, China}}

\newcommand{\Thailand}{\affiliation{National Astronomical Research
    Institute of Thailand, 260 Moo 4, Donkaew, Maerim, Chiang Mai,
    50180, Thailand}}

\newcommand{\UVA}{\affiliation{Department of Astronomy, University of
    Virginia, 530 McCormick Rd, Charlottesville, VA 22904, USA}}

\newcommand{\LJMU}{\affiliation{Astrophysics Research Institute,
    Liverpool John Moores University, 146 Brownlow Hill, Liverpool L3
    5RF, UK}}

\newcommand{\MPIA}{\affiliation{Max-Planck-Institut f\"ur Astrophysik,
    Karl-Schwarzschild Stra{\ss}e 1, 85748 Garching, Germany}}

\newcommand{\JHU}{\affiliation{Physics and Astronomy Department,
    Johns Hopkins University, Baltimore, MD 21218, USA}}

\newcommand{\OSU}{\affiliation{Department of Astronomy, The Ohio State
    University, Columbus, OH, USA}}

\newcommand{\CCAP}{\affiliation{Center for Cosmology and Astroparticle
    Physics, The Ohio State University, Columbus, OH, USA}}

\newcommand{\MIT}{\affiliation{Department of Physics and Kavli Institute for Astrophysics and Space Research, Massachusetts Institute of Technology, 77 Massachusetts Avenue, Cambridge, MA 02139, USA}}

\newcommand{\CPA}{\affiliation{Centre for mathematical Plasma Astrophysics, Department of Mathematics, KU Leuven, Celestijnenlaan 200B, B-3001 Leuven, Belgium}}

\newcommand{\nextinstitute}{\affiliation{Put the institute of the new author here}}


\author{S.~Shiber}
\email{sshiber@fsu.edu}
\FSU

\author{P. Hoeflich}
\FSU

\author{T.~Mera}
\FSU

\author
{E.~Fereidouni}
\FSU

\author
{Z.~Levy}
\FSU

\author
{D. Maci}
\CPA

\author
{C. Ashall}
\IFA

\author
{K. Medler}
\IFA

\author
{J.~M.~DerKacy}
\affiliation{Space Telescope Science Institute, 3700 San Martin Drive, Baltimore, MD 21218-2410, USA}

\author
{E.~Baron}
\PSI
\HS

\author
{M.~Shahbandeh}
\STSci


\author
{C.~M.~Pfeffer}
\GRFP
\IFA


\date{\today}

\begin{abstract}{
Type Ia supernovae (SNe Ia) explosively destroy C/O white dwarfs (WDs) in multiple stellar systems. They produce $\approx 50\%$ of the iron-group elements in the Universe, synthesize electron-capture (EC) elements, drive nuclear physics experiments, and underpin high-precision cosmology. To first order, the outcome is governed by nuclear physics, a property often described as “stellar amnesia.”
Recently, this stellar amnesia has begun to be broken by the nearly universal detection of EC elements with JWST. These elements trace high-density burning, largely ruling out the currently popular He-triggered, sub-$M_{\rm Ch}$ detonation models as the dominant channel. Instead, the ubiquitous presence of EC is shifting back the focus to dynamical and secular mergers, and near-$M_{\rm Ch}$ explosions similar to the deflagration model W7, but in which the nuclear flame undergoes a deflagration-to-detonation transition. The early deflagration phase is especially important because spherical simulations identify the central WD density, and thus the WD mass, as a key parameter governing the explosion.
Here, we present detailed magneto-hydrodynamical simulations. We find that small-scale, pre-existing turbulence expected from the pre-explosion smoldering phase is essential for overcoming the fundamental challenges imposed by the intrinsic 3D physics. This turbulence systematically reduces the production of EC elements by about a factor of two, implying the need for WD central densities closer to those associated with accretion-induced collapse to a neutron star. We also demonstrate the effect of magnetic fields near the saturation field strength and highlight the need for higher-precision EC rates at low $Y_e$.}

\end{abstract}

\maketitle

\section{\label{sec:intro} Introduction}

Supernovae (SNe) are among the most luminous events in the universe.  
The heavy elements in the Galaxy originate from the ejected matter of SNe, where the explosion of white dwarfs (WDs) in thermonuclear SNe (SNe Ia) is responsible for $\gtrsim 50\%$ of the iron-group-elements (IGE) in the Galaxy \cite{iwamoto99}.
They signify the explosive termination of either massive stars or WDs and are therefore important for our understanding of stellar evolution and the chemical distribution in galaxies. 
Despite substantial effort and recent progress, the mechanisms by which stars explode have not been fully explained.

Based on observed luminosities and spectra,
SNe Ia involve the thermonuclear explosion of at
least one carbon–oxygen (C/O) WD \cite{hf60}.
Since nuclear physics governs the progenitor structure, the abundances, and the explosion energies, most explosion scenarios can reproduce the overall light curves (LC) and total-flux spectra, a fact described as “stellar amnesia”\cite{hoeflich2006b}. 
Determining the exact origin of SNe Ia is crucial for their use as extragalactic distance indicators\cite{Riess2017HB}.
Although they serve as ’quasi-standard candles‘\cite{Phillips1993}, growing observational evidence indicates that there is spectral diversity among SNe Ia, which may impact their use in precision cosmology. These observed differences\cite{Jhaetal2019} suggest that SNe Ia may originate from several different
physical processes \cite{WI73,Piersantietal2004,Liu2023,Ruiter2025}.

A critical indicator that can be used to differentiate between the proposed scenarios is the detection of electron-capture (EC) elements \cite{Gerardyetal2007,Galbanyetal2019,Blondinetal2022}. This fact has been greatly boosted since the launch of the James Webb Space Telescope (JWST), which enabled us for the first time to see lines of IGE in the unblended regime of the near- and mid-infrared \cite{DerKacyetal2023,Ashalletal2024,DerKacyetal2024,Kwoketal2025}. Because we can additionally distinguish in this regime between lines of different ionization states, the mass of the elements can be more accurately deduced. Observing at a late time when the SN ejecta become optically thin allows for the inspection of the innermost regions. By this time, the majority of the radioactive $^{56}{\rm Ni}$ has decayed, and any Ni emission is due to stable isotopes such as $^{58}{\rm Ni}$. Radioactive isotopes of other IGEs, such as $^{56}{\rm Co}$ and $^{57}{\rm Fe}$, have longer half-lives and are also products of decays of $^{56}{\rm Ni}$, making their stable abundance more difficult to isolate. 

\begin{figure*}
\centering  
    \includegraphics[scale=0.22, trim={0.0cm 0.0cm 0.0cm 0.0cm}, clip]{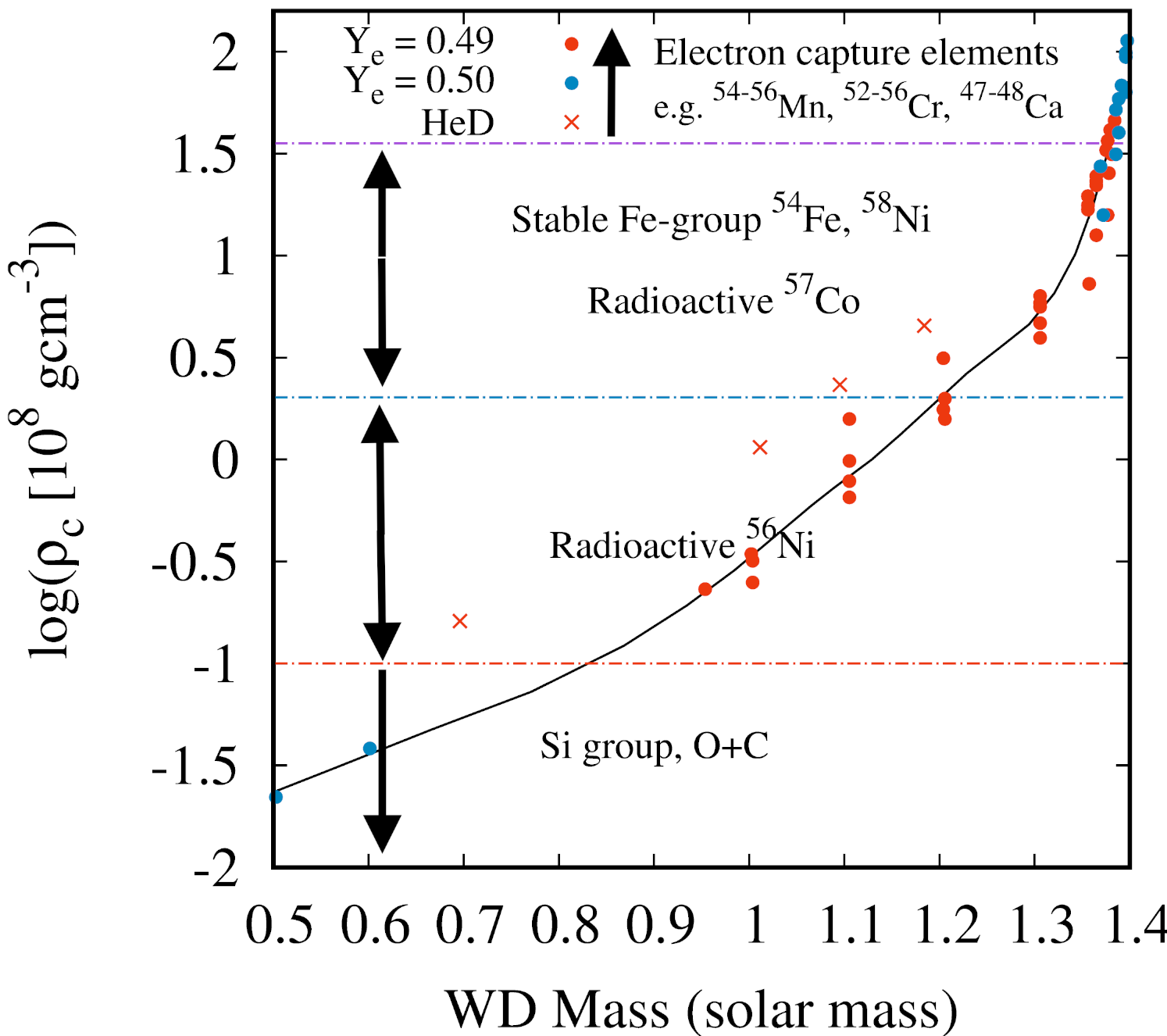}
    \caption{ Hydrostatic WD mass vs.\ central density $\rho_c$ for $Y_e=0.49,0.5$. Adopted from \citealt{2019nuco.conf..187H} with EC ranges added\cite{Brachwitzetal2000,Hoeflichetal2017,Galbanyetal2019}. The solid line corresponds to $\rho_c$ as a function of the total WD mass for C/O=1 and solar metallicity. In sub-$M_{\rm Ch}$ models, the actual burning occurs under slightly higher density, shifting the maximum density of burning upwards relative to near $M_{\rm Ch}$ models. For a given $\rho_c$ (hence $M_{\rm WD}$), only nuclei below the corresponding horizontal line are produced; high-mass explosions yield many electron-capture elements, while $M_{\rm WD}\!<\!1.2\,M_{\rm \odot}$ makes little stable $^{58}$Ni \cite{HoeflichWheeler98,2019nuco.conf..187H}. 
    }
    \label{fig:ec}
\end{figure*}

Although no more than a dozen SNe Ia have published mid-infrared spectra (SNe 2003hv, 2005df, 2006ce, 2014J, 2021aefx, 2022pul, 2022xkq, 2022aaiq, 2023qov, 2024pxl, 2024vjm), all show forbidden lines of $^{58}{\rm Ni}$, which implies that the production of EC elements is a robust feature of SNe Ia. Multidimensional hydrodynamic and spectroscopic modeling of a subset of these objects suggests that $>0.04~M_{\rm \odot}$ of $^{58}$Ni has been formed in each explosion\cite{Gerardyetal2007,DerKacyetal2023,Ashalletal2024,DerKacyetal2024}. As shown in Figure~\ref{fig:ec}, 
The presence of stable $^{58}{\rm Ni}$ points to high-density nuclear burning\cite{Thielemann1986,Brachwitzetal2000} and near-$M_{\rm Ch}$ progenitors ($M_{\rm WD}>1.2~M_{\rm \odot})$. Moreover, stable $^{58}$Ni is mostly observed at low velocities $<3000~{\rm km/s}$, implying that it has been produced in the central regions of the explosion\cite{Ashalletal2024,DerKacyetal2024}.

Nuclear studies have shown\cite{t1,t2,t3,Brachwitzetal2000}
an active EC production on free protons and IGE during the early burning stage of a thermonuclear SN Ia explosion. The EC reduces the average electron number to nuclei, $Y_{\rm e}$, from its original value, leading to the creation of nuclei in the range between $N=Z$ and stability ($N>Z$). Unlike the conditions in core collapse SNe \cite{Martinezetal2000}, $\beta$-decay does not play a dominant role, and the most neutron-rich nuclei encountered include $^{48}$Ca, $^{50}$Ti, $^{54}$Cr, and $^{58}$Fe, while $^{54}$Fe and $^{58}$Ni are produced for more moderate $Y_{\rm e}$ values (as depicted in Figure~\ref{fig:ec}). 
The EC rates have been modified over the years, thanks to advances in both computational methods and nuclear experiments \cite{Thielemann1986,Brachwitzetal2000,Thielemannetal2004,Langanke04}. Specifically, the weak rates compilation of \citealt{LangankeMartinez2001} significantly reduced the EC capture rates on nuclei compared to previous compilations \cite{FFN82a}. 

In sub-$M_{\rm Ch}$ scenarios, like He-detonation models \cite{wwt80,Nomoto1982_I,livne1990,Woosley94,hk96,Kromeretal2010,Sim10,WoosleyKasen2011,Shen2015,Tanikawa2018,Glasner2018,2019arXiv190310960T}, the helium shell ignites and sends a detonation wave that triggers the detonation of the entire C/O WD. However, in these models, the burning occurs at relatively lower densities, and therefore, they do not produce a considerable amount of EC elements. The short duration of burning, which propagates as a detonation wave at supersonic speed, additionally limits EC production. Near-$M_{\rm Ch}$ models, in which mass is gained either by a merger with another WD or by accretion, can produce a higher amount of EC elements. Current spherical delayed detonations (DDT) models well reproduce LC and spectra from the optical to mid-IR \cite{Hoeflichetal2017,Ashalletal2024}. In these models, burning starts at the center as a subsonic deflagration that propagates outward isotropically until it reaches a transition density at which a detonation is initiated \cite{Khokhlov1991}. Since the density structure is spherical in DDT models, it is more consistent with polarimetry studies \cite{Patatetal2012} than with mergers (either violent or secular), which should result in an aspherical structure. 

However, key physical processes in the DDT explosion mechanism, such as buoyancy, turbulence, and the deflagration–detonation transition, introduce asphericities, requiring three-dimensional full-star simulations for a self-consistent approach to the problem. Inconsistent with spherical models, multidimensional studies of the deflagration phase have found that as the deflagration propagates, plumes are formed as a result of a buoyancy force \cite{Gamezoetal2003, Gamezoetal2004, Gamezo_etal_2005,seitenzahl13,Lachetal2022}, leaving a large fraction of unburned matter near the star center. EC elements are produced mainly in these rising plumes, reducing their amounts and shifting their distributions to higher velocities compared to spherical models\cite{Pakmor2024}. In 3D, the WD only moderately expands before the detonation, creating only low yields of intermediate-mass elements (IME), and the resulting spectra do not match the observations.

In a recent study, \citealt{Shiberetal2026} has shown that small-scale turbulence might offer a remedy to this problem. The preexisting small-scale and strong turbulence drags the deflagration between the plumes during the early deflagration phase, before and slightly after Rayleigh-Taylor (RT) instabilities start to develop, burning these pockets of unburned matter and causing a more complete and spherical burning. This preexisting turbulence, with velocities of up to 500 km/s and eddies of up to $\approx 50$ km, is expected to form in the pre-explosive burning during the smoldering phase \cite{HoeflichStein2002,Zingaleetal2011}. However, only the early phase has been simulated in \citealt{Shiberetal2026}, and the question of whether this efficient burning persists for a longer duration still remains. In addition, the production of EC elements in these cases needs to be carefully examined and compared with observations. This is the purpose of the current study.

\begin{figure}
\centering  
    \includegraphics[scale=0.4, trim={0.0cm 0.0cm 0.0cm 0.0cm}, clip]{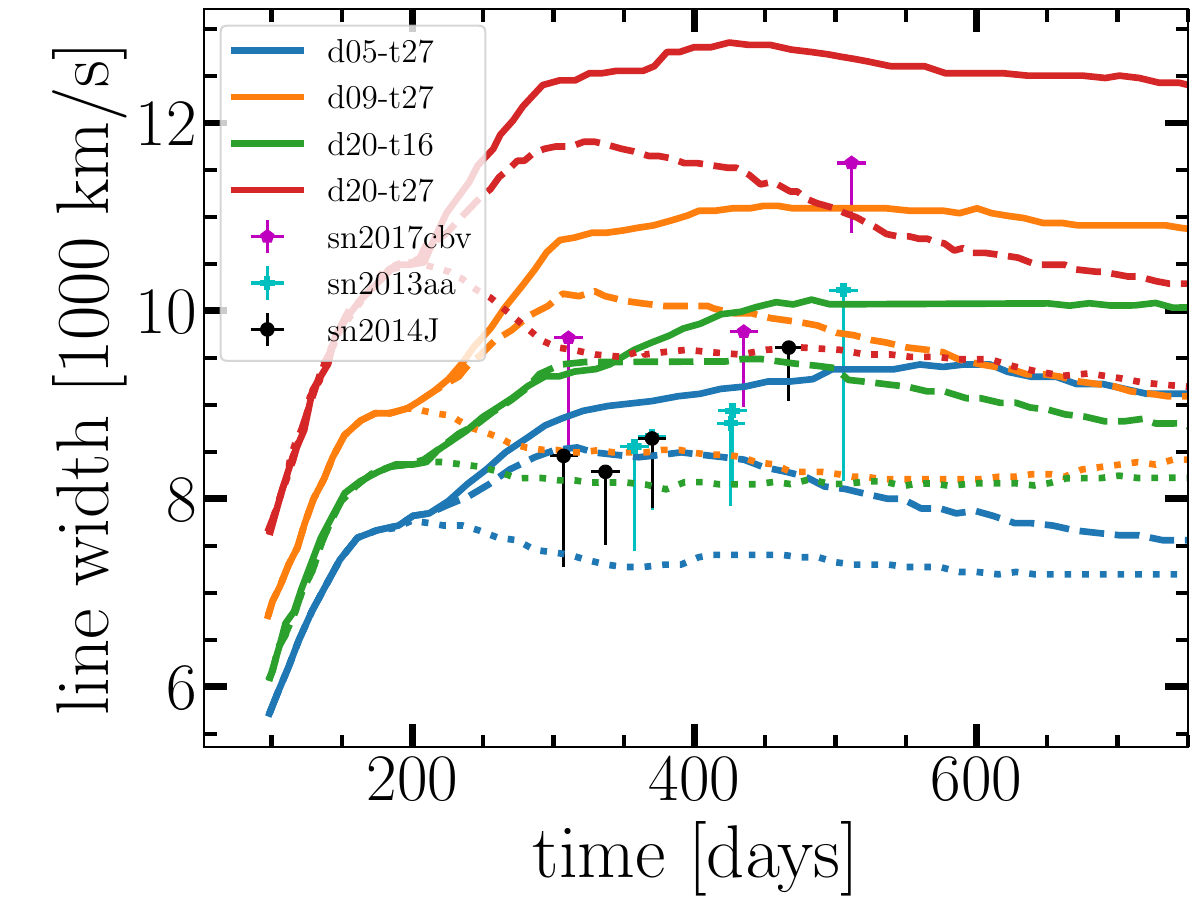}
    \caption{  { Time evolution of the width of the [Fe~II ] 1.644 $\mu$m feature as observed in SN 2013aa, SN 2017cbv, and SN 2014J \cite{Kumaretal2026,Diamondetal2018} compared to values from simulations based on the spherical delayed-detonation scenario \cite{Diamondetal2015,Diamondetal2018,Hristovetal2021}. For the observations, error bars reflect uncertainties in the estimates of the underlying continuum. The simulations are characterized by the central density of the WD, the size of the initial magnetic field, and the transition density at which the deflagration to detonation occurred. 
    Each color represents different simulation named dXX-tYY with XX being the central density (in $\times10^{8}~{\rm g~cm^{-3}}$) and delayed-detonation transition at YY (in $\times10^{6}~{\rm g~cm^{-3}}$), while different line styles represent different initial magnetic fields of $10^3$ (dotted), $10^6$ (dashed), and $10^{9...13}$ (solid) G. The line width depends on both the EC in a high-density core and the initial magnetic field. The initial rise is caused by the envelope becoming optically thin and a larger velocity range being seen, followed by a phase of positron escape from the outer, high-velocity layers. With increasing B, the positrons remain increasingly trapped. As a result, the curves peak later, and they become flatter \cite{Hoeflichetal2004,Penneyetal2014}.  
    All three SNe Ia show sustained increase in line widths as late as 500 days, indicating ultra-high, initial magnetic fields. }}
    \label{fig:widths_with_b}
\end{figure}

 {Furthermore, it has been shown that in order to explain late-time LCs \cite{Hristovetal2021} and line widths of the unblended [Fe~II] 1.644 $\mu$m line 
\cite{Hoeflichetal2004,Diamondetal2018}, high magnetic fields in excess of $\approx 10^{6...7}~{\rm G}$ must exist in the progenitor WD. Recently, new spectral observations of three SNe~Ia became available out to 500+ days  \cite{Kumaretal2026}, stretching the current limits on magnetic fields $B$. Figure~\ref{fig:widths_with_b} presents our analysis of the new data along with DDT models that include magnetic fields with different strengths \cite{Diamondetal2015,Diamondetal2018,Hristovetal2021}.} 
 {The radioactive decay $^{56}$Co~$\rightarrow$~$^{56}$Fe has two main channels: electron capture resulting in an excited $^{56}$Fe nucleus that emits $\gamma$-rays when going to the ground state; and $\beta^+$ decay resulting in fast positrons. At early times, $\gamma$-rays dominate the energy input into the SN envelope, but by day 200, they mostly escape, and positrons increasingly dominate the energy input \cite{Hoeflichetal2004,Penneyetal2014}. 
The strength of the surrounding magnetic field determines the duration for which positrons stay trapped in the Fe region, prolonging the increase of the line width beyond 500+ days. In the observations, lines still widen at +300 and +500 days, strongly supporting the existence of ultra-high initial magnetic fields. A similar evolution is observed in three different well-observed SNe Ia, suggesting that high B fields may be common.}

These fields are stronger than typical fields found in isolated WDs \cite{liebert03}. A potential amplification process is through a large-scale dynamo mechanism during the accretion phase that would lead to a large-scale dipole field. Another potential dynamo mechanism can act during the smoldering phase, where small-scale dynamos are produced by convection to form a small-scale unstructured magnetic field \cite{Brandenburg05,Beresnyak12,Tayler73,Acheson78,Hawley96,Spruit02,braithwaite09,Duez10a,Duez10c}. 
In \citealt{Shiberetal2026}, turbulent magnetic fields of less than a percent of the saturation strength have been found to affect the burning only slightly. In the current study, we extend the work by running a simulation with a stronger magnetic field and find that strong magnetic fields dominate over the turbulence. We also simulated magnetic fields with large-scale dipole morphology.

As mentioned, the main aim of this paper is to present a coherent, up-to-date picture of EC production in SNe Ia and, in particular, to investigate the influence of preexisting fields on the production of EC elements. We analyze a set of ideal magneto-hydrodynamic (MHD) 3D simulations with different initial magnetic and turbulent velocity fields and compare them to spherical DDT models that well reproduce observations. We describe our numerical tools and simulation setup in Section~\ref{sec:setup}. Section~\ref{sec:results} presents our results, while in Section~\ref{sec:conclusions} we conclude.

\section{\label{sec:setup} Numerical setup}

We use the \texttt{FLASH}\cite{Fryxelletal2000} multi-physics code, version 4.8, to carry out 3D MHD simulations of the early deflagration phase in near-$M_{\rm Ch}$ WDs. We evolved the WD from a $7~M_{\rm \odot}$ zero-age main-sequence star of solar metallicity\cite{1989ApJ...344..239C,1999ARA&A..37..239B,2001ApJ...557..279D}. The core is carbon depleted as a result of stellar central helium burning.
The WD accretes hydrogen (and helium) from a companion, reaches a mass of $M_{\rm WD}=1.35 M_\odot$, and contracts to a radius of $\simeq 2,400~{\rm km}$ \cite{2001ApJ...557..279D,Hoeflichetal2017}. The accretion rate is adjusted to lead to a thermodynamic runaway at a central density of $\simeq 10^9~{\rm g~cm^{-3}}$ \cite{SugimotoNomoto1980,Nomoto1982_I,Hoeflich_2002,2019nuco.conf..187H,Hristovetal2021}.

 {We map the WD structure to a 3D, non-uniform, adaptive mesh refinement (AMR) grid of \texttt{FLASH}, with cartesian coordinates and dimensions of $\left(5\times10^3~{\rm km}\right)^3$.
The base AMR resolution in all our simulations consists of $\left[32\times32\times32\right]$ cells, corresponding to a refinement level $l=3$ and a maximum cell size of $dx_{\rm max}=165~{\rm km}$. Each additional refinement increases the resolution by a factor of $2$ in the refined region. The maximum refinement level, $l_{\rm max}$, varies between simulations and ranges from 7 to 10, translating into a minimal cell size of $dx_{\rm min}=10~{\rm km}$ to $1.2~{\rm km}$, respectively. The maximum level of refinement for each simulation is listed in the rightmost column of Table~\ref{tab:prop}.} 
The grid is logarithmically refined from the center out, while enforcing maximum refinement in the inner region.
 {For illustration, a view of the computational grid of simulation turb4b, including a zoom-in over the innermost region, can be found in Figure 1 of \citealt{Shiberetal2026} (referred to there as simulation D40V170B12).}

The gravity is computed using \texttt{FLASH}'s Poisson multipole solver with a highest multipole order of 16 and isolated boundary conditions. To solve the MHD equations, we employ \texttt{FLASH}'s unsplit staggered mesh solver, assuming ideal MHD, with a zero gradient boundary condition. We utilize a hybrid Riemann solver, which combines the Roe solver for high accuracy and HLLD for stability. A Helmholtz equation of state, which includes contributions from degenerate electrons–positrons, thermal ions, radiation, and Coulomb corrections, is used \citep{TimmesSwesty2000}.  {In Appendix~\ref{appendix:mhd-eqs}, we explicitly list the full set of MHD equations solved by \texttt{FLASH}}.  

The deflagration propagates according to the advection-diffusion-reaction (ADR) scheme \citep{Khokhlov1995} with a sharpened Kolmogorov Petrovski Piskunov (sKPP) reaction term \citep{Vladimirova2006}, where default values for $\epsilon_0,~\epsilon_1,~f_b$,~and~$b$ were used (see \citealt{Townsley2007}). We assume nuclear statistical equilibrium (NSE) for densities $\rho>2\times 10^{7}{\rm ~g~cm^{-3}}$, nuclear statistical quasi-equilibrium (NSQE) for $2\times 10^{7}{\rm ~g~cm^{-3}}\ge \rho > 10^{6}{\rm ~g~cm^{-3}}$, and carbon burning for $10^{6}{\rm ~g~cm^{-3}}\ge \rho$. Since we focus on the early phase of the deflagration, where the front remains in high-density regions, the main burning is through NSE, while NSQE and carbon burning contribute only marginally. We track the burning with a scalar field $\phi$, where zero means unburned fuel and one means fully burned ashes. 

The rate of NSE nuclear (specific) energy is: $\dot{E}^{\rm NSE}_{\rm nuc}=\left(\dot{\phi} /m_p\right)\times\left[\tilde{b}_{\rm Ni} - (1 - {\rm X}_{\rm C})\tilde{b}_{\rm O} - {\rm X}_{\rm C}\tilde{b}_{\rm C}\right]$, where $\tilde{b}_{\rm Ni}$, $\tilde{b}_{\rm O}$, and $\tilde{b}_{\rm C}$ are the binding energies per nucleon of $^{56}{\rm Ni}$, $^{16}{\rm O}$, and $^{12}{\rm C}$, respectively. Since  the WD core is carbon depleted (${\rm X}_{\rm C}=0.28$ for $\rho\ge2\times 10^{8}$), we get $\dot{E}_{\rm nuc}^{\rm NSE}= 7.2\times10^{17} \dot{\phi}~{\rm erg~g^{-1}}$. This might underestimate the generated nuclear energy when the deflagration propagates outward to lower-density regions.
For NSQE and carbon burning, we assume the consumption of an even C/O mixture to Si/S and O/Mg/Si, respectively, i.e. $Q^{\rm NSQE}= 6.1\times10^{17}~{\rm erg~g^{-1}}$ and $Q^{\rm C}= 2.5\times10^{17}~{\rm erg~g^{-1}}$. We do not change the composition itself according to the burning in the MHD code. This might have a small effect on the temperature. 

 {We emphasize that during the early deflagration phase, the thickness of the deflagration front (flame) is many orders of magnitude smaller than the simulation cell size. To incorporate subgrid effects on the deflagration propagation speed, $v_{\rm burn}$, a subgrid scheme is usually employed. Such a scheme commonly adopts
$v_{\rm burn}=max(v_{\rm laminar},v_{\rm turb,sub})$, where $v_{\rm turb,sub}$ refers to the small-scale turbulence below the grid resolution and $v_{\rm laminar}$ is the laminar flame speed, which typically equals $\approx 100~{\rm km/s }$ in the central region \cite{khokh01,Gamezoetal2003}. This is to be differentiated from the relevant scales of the pre-existing turbulence, $v_{\rm turb}$, which we include in our simulations as initial conditions. These pre-existing turbulent fields are the result of the prior smoldering phase \cite{HoeflichStein2002,Zingaleetal2011}, and the large scales relevant for the mixing of burned and unburned matter are well resolved within our grid, with at least 30 cells across the turbulence diffusion radius (see below). }

 {Other subgrid models of the flame speed include induced turbulent flame speed closures \citep{NiemeyerHillebrandt1995,Schmidtetal2006} and other sophisticated models \citep{Pateletal2026}. Alternatively, $v_{\rm burn}$ is treated as a constant ranging from $50-200~{\rm km/s}$, e.g.,  \citealt{Maetal2013, Hristovetal2021}. In their study \cite{Maetal2013}, they found only small differences between a fixed flame speed of $100~{\rm km/s}$ and $200~{\rm km/s}$. Thus, in this study, a constant $v_{\rm burn}$ of $200~{\rm km/s}$ has been adopted.   
This higher velocity was chosen to include both the non-resolved scales of the pre-existing turbulent field and the flame-induced turbulence later on.} 

 {Note that all the prescriptions mentioned above assume a static WD essentially as initial conditions. However, WDs are not stationary: 1) The pre-existing $v_{\rm turb}$ eddies will decay, adding sub-cell turbulence from the sub-cell scale to the flame thickness; 2) the rotation of the WD will induce large-scale circulations \cite{Hachisu1986b} and cascading instabilities that add small-scale motion; 3) G-mode instabilities have been widely observed in WDs \cite{Wingetetal1994}, adding another source of motion; 4) the transition from the smoldering phase to the explosive phase is not well understood but is likely related to instabilities on small scales; and 5) burning instabilities at the surface prior to the explosion can propagate inward to the center. All of these processes will add to the sub-grid turbulence, $v_{\rm turb,sub}$ and to $v_{\rm burn}$. However, the total contribution will be influenced by the abundances, magnetic fields, viscosity, etc. Therefore, for the purposes of this study, we favor a constant burning velocity as the simplest approach to avoid the uncertainty in the specific processes involved and use $200~{\rm km/s}$ as a fiducial value. Furthermore, in tests done with lower-resolution models, we simulated a slower speed of $100~{\rm km/s}$ and found qualitatively similar behavior, confirming the results of \citealt{Maetal2013}.}

For the initialization of the turbulence, we use the BxC toolkit, designed to generate fully customizable synthetic turbulent 3D magnetic fields \cite{Durriveetal2022,Macietal2024}. We set the turbulence properties to closely follow the physical conditions expected from the smoldering phase \cite{HS02}; i.e., we set the diffusion radius,  {corresponding to the high wavenumber end of the Kolmogorov spectrum}, to $\approx 50$ km and $v_{{\rm rms}} $ up to $\approx 200 $ km/s. The maximally refined innermost $\left(400~{\rm km}\right)^3$ regions well resolved the turbulence, while exterior to this box, the turbulent pattern is duplicated.  {We note that the developed velocity field during the smoldering phase depends on the initial abundance structure of the WD. In \citealt{HS02}, the abundance gradient between low C and $C\approx 0.5 $ constrains the turbulent region to the center. In contrast, simulations without this constraint, e.g., \citealt{Zingaleetal2011}, result in a slower velocity field with $v_{\rm rms} \approx 50~{\rm km/s}$, likely because the turbulence is distributed over the entire WD rather than simmering only in the smaller confined central region.} 

\citealt{Hristovetal2021} found that an initial large-scale dipole magnetic field has become turbulent during the deflagration phase. Consequently, the magnetic field would become turbulent even before the deflagration starts; therefore, we assumed the morphology of the turbulent $B$ field as the background turbulence. 
We note, though, that the initial turbulent field may not resemble the background turbulence if the magnetic pressure becomes comparable to the gas pressure.
Therefore, we constrain the magnetic field strength in our simulation to be less than 10\% of the gas pressure.

For the EC production, nucleosynthesis is performed in a post-processing manner by analyzing tracer particles. We initialized a subset of our simulations with passive tracer particles. The particles are initially positioned randomly, with particle density being proportional to the grid gas density, i.e., a higher number of particles are positioned within regions of higher mass. This ensures that each particle has a mass equal to the total mass in the grid divided by the number of particles. The exact number of particles varies between 7,500 and 8,500 depending on the simulation resolution. Particle positions are advanced in time according to the Two-Stage Runge-Kutta scheme (also known as the Heun’s method), while the interpolation of velocity, density, and temperature from the grid to the location of the particles uses a quadratic mapping scheme. By analyzing each particle trajectory separately, that is $\rho_i(t),~T_i(t)$, in a nuclear reaction module that includes EC rates\cite{Hoeflich2009}, we can infer the EC element yields and compare them to the spherical reference model yields (for details, see Section~\ref{subsec:ec}).

\begin{figure*}
\centering  
    \includegraphics[scale=0.56, trim={0 5.4cm 0 5.1cm}, clip]{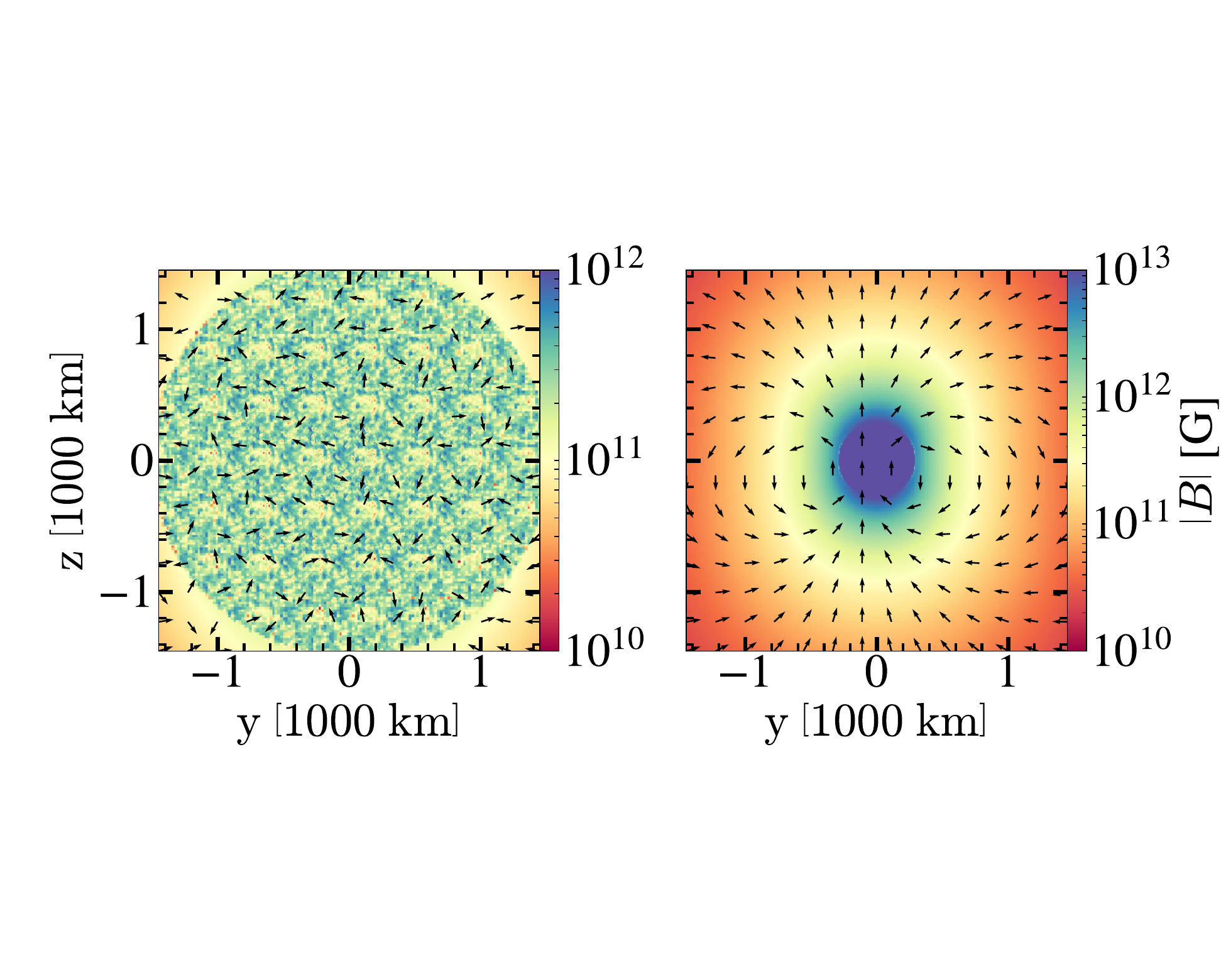}
    \caption{ Initial magnetic field at the yz plane in two simulations representing turbulent (left) and dipole (right) magnetic fields. Left: simulation turb4b, corresponding to smallest eddies size of roughly $40~{\rm km}$ and a rms magnetic field of $3\times10^{11}~{\rm G}$. Right: simulation dip3a, corresponding to a large-scale dipole magnetic field with a strength of $10^{12}~{\rm G}$ at $r=(z=0,~R_{\rm cyl}=500~{\rm km})$. To avoid ultra-high values at the center, we constrain the magnitude of the dipole field to $10^{13}~{\rm G}$.  {Note that the refined structure of the AMR grid of simulation turb4b (here on left), including a zoom-in view on the innermost region, can be found in Figure 1 of \citealt{Shiberetal2026} (being referred there as simulation D40V170B12), illustrating that we well resolved the relevant scales of the preexisting turbulent fields. }}
    \label{fig:B_setup}
\end{figure*}

We list the properties of our simulations in Table~\ref{tab:prop}, while in Figure~\ref{fig:B_setup} we present the initial magnetic field morphology in two specific simulations: turb4b (left) and dip3a (right), representing small-scale turbulent and large-scale dipole magnetic fields, respectively. The magnetic field magnitude is shown on the meridional yz plane together with arrows indicating the direction of the magnetic field. 
To avoid ultra high values at the center, we constrain the magnitude of the dipole field to $10^{13}~{\rm G}$.
\begin{table*}
\centering
\begin{tabular}{l|lllll}
Simulation & $B$ & $r_D$ & $v_{\rm rms}$ & $B$ strength & $l_{\rm max}$  \\
 & morphology & [km] & [km/s] & [G] &   \\
\hline
wob1a     & --        & --  & --  & --                & 7   \\
wob1b     & --        & --  & -- & --                 & 10  \\ 
wob2      & --        & 300 & 30 & --                 & 7   \\
wob3      & --        & 40  & 20  & --                & 10  \\
\hline
dip1   & dipole    & --  & --  & $10^{6}$          & 7   \\
dip2a  & dipole    & --  & --  & $10^{9}$          & 7   \\
dip2b  & dipole    & 300 & 30  & $10^{9}$          & 7   \\
dip3a  & dipole    & --  & --  & $10^{12}$         & 7   \\
dip3b  & dipole    & 300 & 30  & $10^{12}$         & 7   \\
\hline
turb1  & turbulent & 80  & 170  & $3\times10^{11}$ & 9   \\
turb2  & turbulent & 40  & 20   & $3\times10^{11}$ & 10  \\
turb3  & turbulent & 40  & 50   & $3\times10^{8}$  & 10  \\
turb4a & turbulent & 40  & 170  & $3\times10^{5}$  & 10  \\
turb4b & turbulent & 40  & 170  & $3\times10^{11}$ & 10  \\
turb4c & turbulent & 40  & 170  & $5\times10^{12}$ & 10  \\
\end{tabular}
\caption{Properties of our simulations. 
 {$r_D$ is diffusion radius of the turbulence, corresponding to the high wavenumber end of the Kolmogorov spectrum \cite{Durriveetal2022,Macietal2024}}; $v_{\rm rms}$ is the turbulence velocity rms; $B$ strength is the magnetic field rms for turbulent magnetic fields and the magnetic field strength at a cylindrical radius of $500~{\rm km}$ along the equatorial plane for dipole magnetic fields. $l_{\rm max}$ is the maximal grid refinement level. }
\label{tab:prop}
\end{table*}

 {We stress that turbulent and dipole magnetic fields have different scales and a different physical origin, namely
the smoldering phase or the large-scale circulation due to rotation induced by accretion, respectively (see Section~\ref{sec:intro}). Therefore, we include in our set of simulations with turbulent magnetic fields (i.e., our turb series) turbulence with properties expected from the smoldering phase. Our set of simulations with dipole magnetic fields (our dip series) corresponds to a large-scale amplification process that is not related to the smoldering phase; therefore, we include preexisting turbulence with larger scales or do not include preexisting turbulence at all in this set of simulations. All of the simulations assume an ab initio magnetic field configuration and do not follow its creation.}

 { We used the BxC tool to produce a monopole-free initial stationary MHD turbulence generated by non-explosive burning over long time scales. We note that the spatial resolution during the hydrodynamical simulation is insufficient to maintain or develop the initial Kolmogorov spectrum from pre-existing large scales. However,
the turnaround time for the largest eddies, $\approx 1~{\rm s}$, is larger than the characteristic dynamical time scale. Therefore, even at a high resolution, a Kolmogorov spectrum for the pre-existing turbulence cannot develop. Consequently, the large scales of the pre-existing turbulence will be preserved until large-scale hydrodynamical effects take over. Importantly, these large scales drive the abundance mixing.}

After mapping the WD structure to the grid, we damp the velocities for $0.1~{\rm s}$ to allow the star structure to relax before we ignite it. In addition, to avoid inflow from the surrounding low-density medium, we enforce a zero velocity region outside a radius of $1,500~{\rm km}$ by setting the velocities to zero at every time-step in this outer region. However, in order to allow for the WD expansion, we relax this constraint once the deflagration front is close to this outer region. The exact moment in which we relax this constraint has only a minor effect on our results. The reason is that once the deflagration reaches this outer region, it is already well developed, and the WD expansion has already set in. Moreover, by this time, the expanding WD carries enough momentum and thus prevents inflow from the surrounding vacuum, making the enforcement of zero velocity in this outer region no longer required.

\section{\label{sec:results} Results}

In this section, we present the results of our simulations listed in Table~\ref{tab:prop}. In all simulations, the WD is ignited in the center with an ignition radius of $100~{\rm km}$. In Section~\ref{subsec:mhd}, we focus on the influence of preexisting turbulence and $B$ fields on the propagation of the deflagration and on the generation of nuclear energy, while in Section~\ref{subsec:ec}, we focus on the influence of the fields on the synthesis of EC elements.

\subsection{\label{subsec:mhd} Magneto-Hydrodynamics}

\begin{figure*}
\centering  
   \includegraphics[scale=0.6, trim={1.0cm 7.1cm 0cm 4.5cm}, clip]{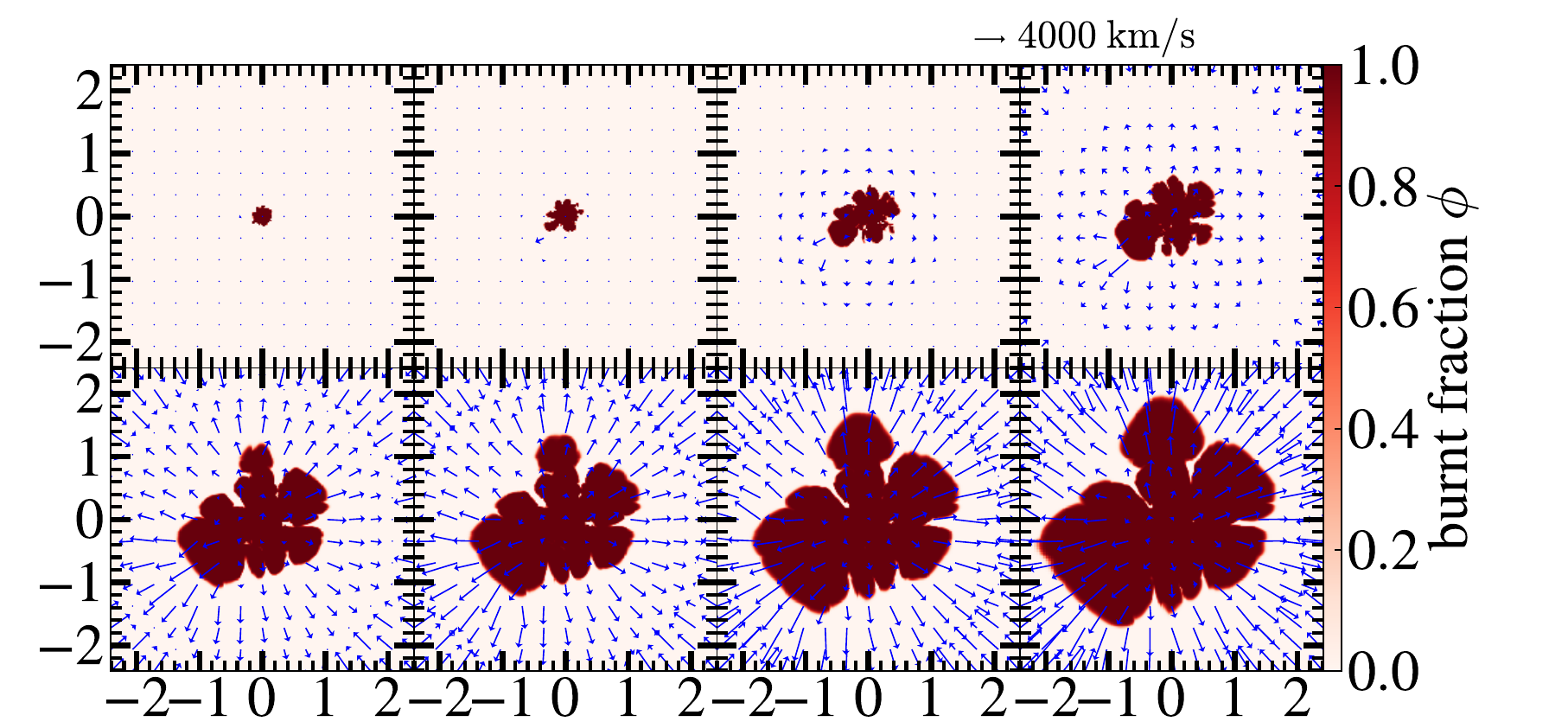}
    \caption{ Time evolution of simulation turb4b. Axes are in units of $1000~{\rm km}$. Blue arrows correspond to velocity vectors with the scale as appears above the figure. Small-scale (turbulence diffusion radius of $r_{\rm D}=40~{\rm km}$) and strong turbulence ($v_{\rm rms} =170~{\rm km/s}$) in this simulation efficiently burns away the pockets and results in more complete burning.}
    \label{fig:turb_evol}
\end{figure*}

\begin{figure*}
\centering  
   \includegraphics[scale=0.6, trim={1.0cm 5.8cm 0cm 4.5cm}, clip]{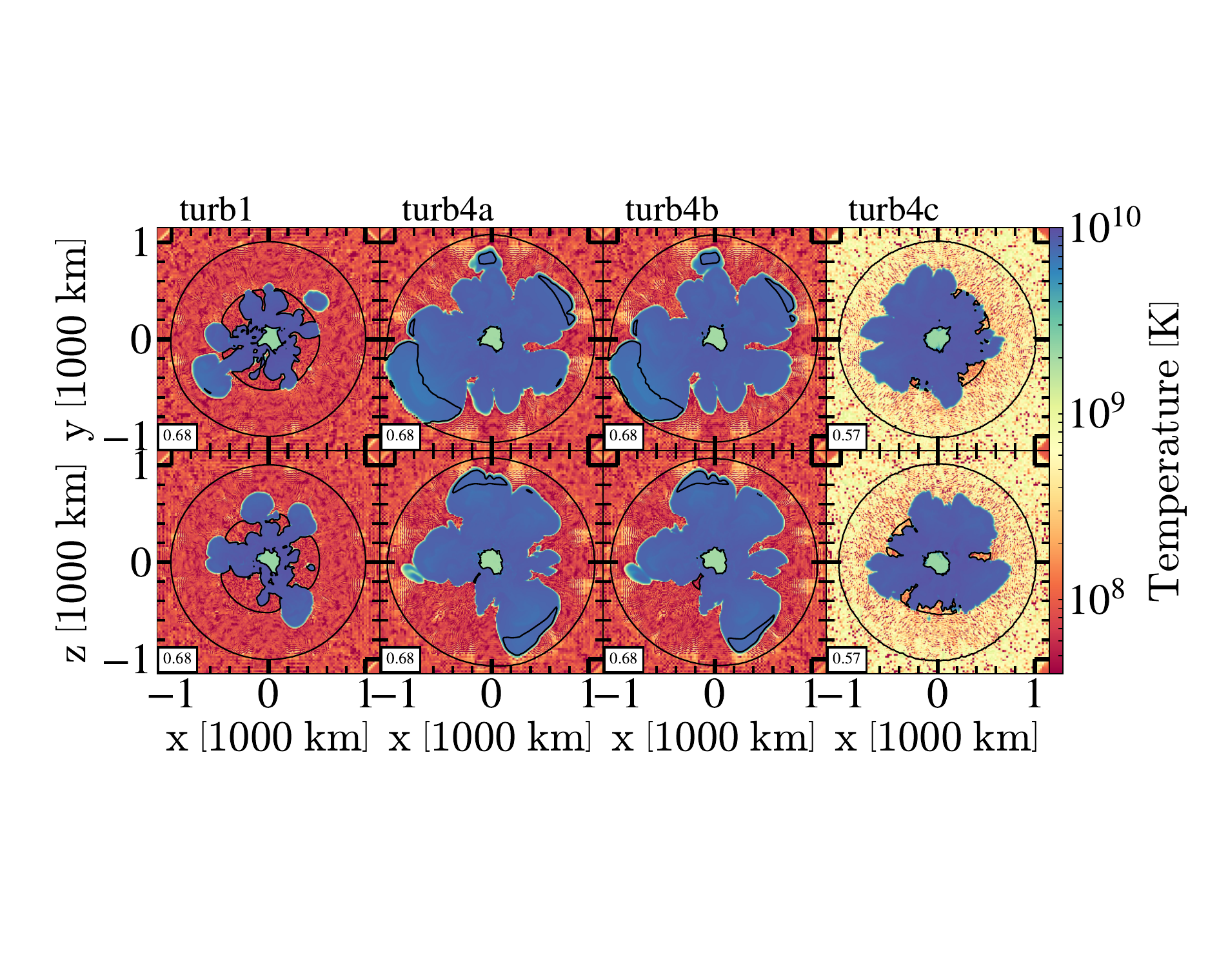}
    \caption{ The effect of small-scale turbulent fields. Shown are temperature slices along the xy plane (equatorial plane; upper panels) and along the xz plane (meridional plane; lower panels) at $t=0.58~{\rm s}$ ($0.47~{\rm s}$ for simulation turb4c).}
    \label{fig:turb_temp}
\end{figure*}

Figures~\ref{fig:turb_evol}~and~\ref{fig:turb_temp} demonstrate that preexisting turbulent fields significantly impact burning. Figure~\ref{fig:turb_evol} presents the evolution of the deflagration front in simulation turb4b, where the burned fraction $\phi$ refers to the mass fraction of burned matter in each cell. Figure~\ref{fig:turb_temp} shows temperature maps in two different planes at an intermediate time. Turbulence with properties expected from the smoldering phase, namely a small diffusion radius $r_{\rm D}=40~{\rm km}$ and strong turbulence $v_{\rm rms} = 170~{\rm km}$, as in simulation turb4b, leaves no pockets of unburnt material near the center and efficiently burns the fuel within the deflagration radius. 

\begin{figure*}
\centering  
   \includegraphics[scale=0.6, trim={1.0cm 5.4cm 0cm 4.5cm}, clip]{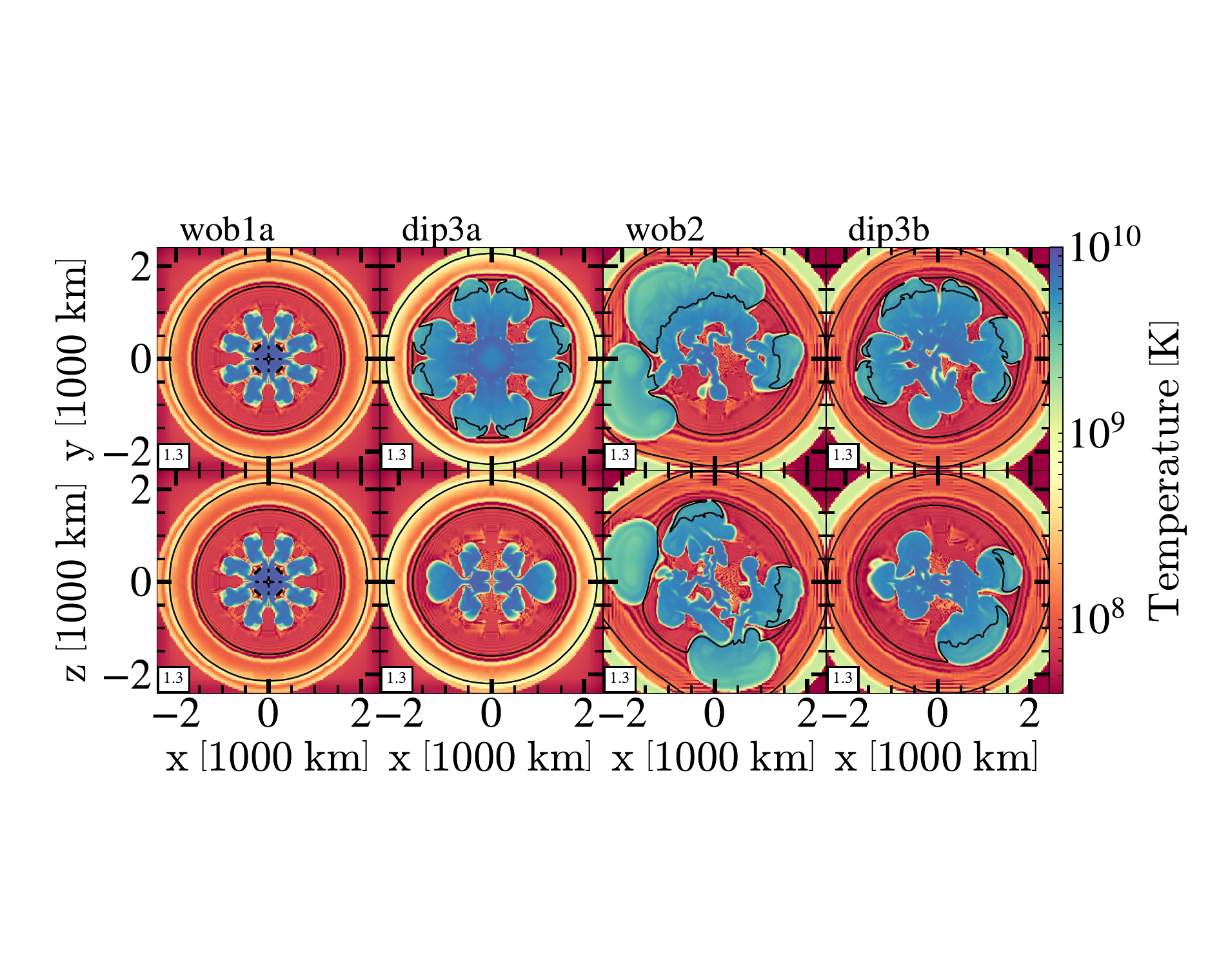}
    \caption{ The effect of a large-scale dipole magnetic field. Shown are temperature slices along the xy plane (equatorial plane; upper panels) and along the xz plane (meridional plane; lower panels) at $t=1.2~{\rm s}$.}
    \label{fig:dip_temp}
\end{figure*}

When the turbulence scale is larger $r_{\rm D}=80~{\rm km}$, as in simulation turb1, plumes of burned material are formed that rise due to buoyancy forces, leaving a large space of unburned fuel between them (first column of Figure~\ref{fig:turb_temp}). The pocket size is even larger for larger-scale turbulence or for cases without turbulence, e.g., simulations wob2 and wob1a, respectively (see Figure~\ref{fig:dip_temp} and the discussion below). These conclusions have been previously demonstrated \cite{Shiberetal2026}, but here we additionally show that the effective burning, which is close to spherical burning, persists for a longer time.

Turbulent $B$ fields with strengths that are less than a percentage of the equi-partition values have only minor effects and mildly impact small-scale structures. Therefore, the burning in simulations turb4a and turb4b, which differ only in their initial magnetic field strengths of $\approx10^{5}~{\rm G}$ and $\approx10^{11}~{\rm G}$, respectively, is very similar (see Figure~\ref{fig:turb_temp} second and third columns). However, stronger turbulent magnetic fields, like in simulation turb4c ($\sim$7\% of the equi-partition field), further mix the unburned and burned matter, resulting in fewer and smaller pockets and more complete burning (fourth column).

In the large-scale dipole magnetic field simulations, we find that they fail to mix unburned and burned matter on a small scale; therefore, large unburned pockets are still formed. However, we find a squeezing effect in response to a strong dipole field, where the deflagration is less developed along the magnetic dipole direction and overdeveloped in the perpendicular direction. 

To illustrate this effect, we show in Figure~\ref{fig:dip_temp} temperature maps in two different planes: the xy plane perpendicular to the dipole (upper panels) and the xz plane along the dipole (lower panels). The strong dipole field of $\approx 10^{12}~{\rm G}$ in simulations dip3a and dip3b inhibits the development of Rayleigh-Taylor instabilities along the $z$ direction, and the burning front is squeezed toward the perpendicular direction. Also note the effect of large-scale turbulent fields of $r_{\rm D}=300~{\rm km}$ in simulations wob2 and dip3b, which still leave a large fraction of the inner region unburned. Most of the EC takes place in the burning plumes; therefore, while not significant to the efficiency of burning and the EC element yields, strong dipole magnetic fields substantially affect the distribution of the EC elements. This effect on the EC distribution should be studied in more detail in the future by simulating the further evolution of the deflagration phase.

The efficiency of burning compared to the spherical case is well demonstrated through an angle-averaged value of the burned fraction within a volume, the filling factor $f$. Filling factor values close to 1 imply spherical burning, while small values imply inefficient burning with large pockets of unburned fuel.

\begin{figure*}
\centering  
    \includegraphics[scale=0.36]{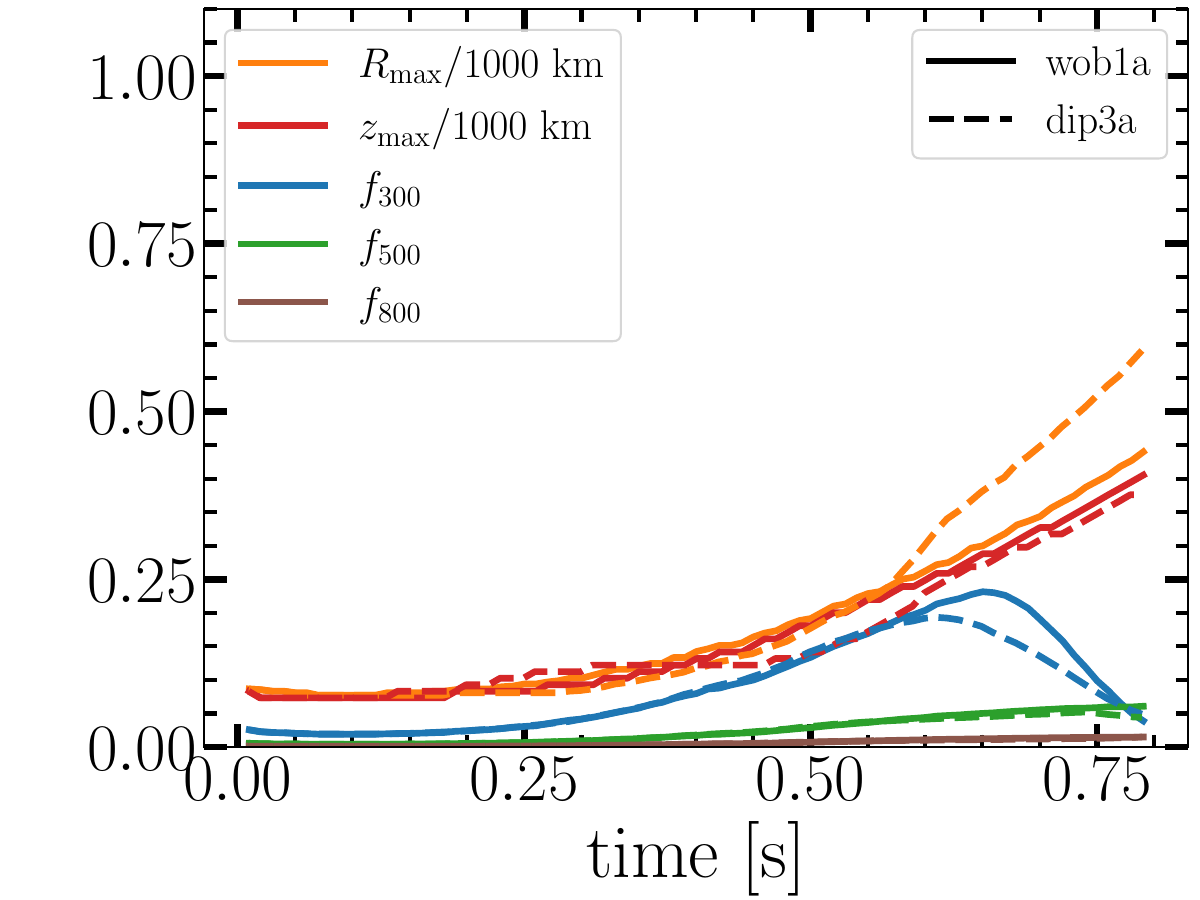}   
    \includegraphics[scale=0.36]{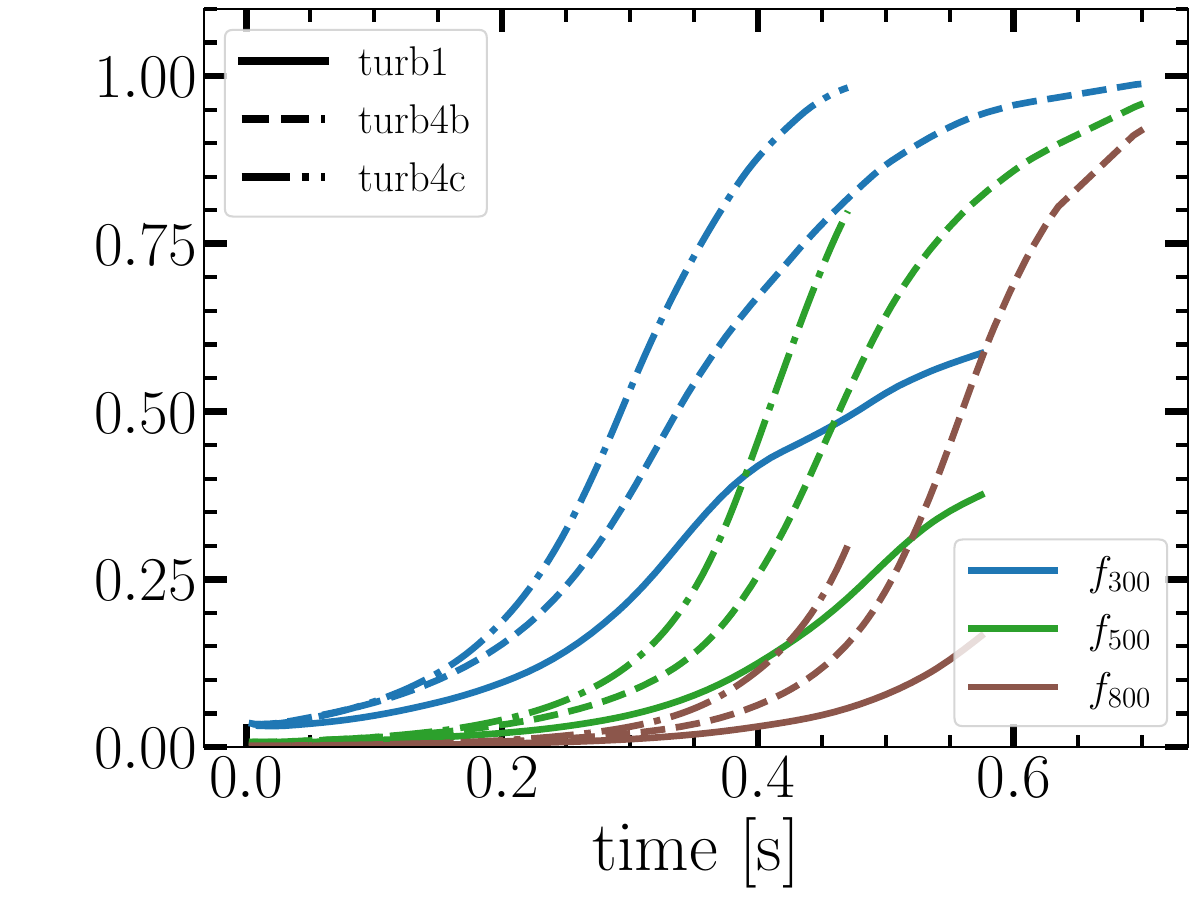}
        \caption{Filling factors of burned and unburned material at 300, 500 and 800 km for various models, and maximum distance from the center $R_{\rm max}$ and along the symmetry axis $z_{\rm max}$ are given as a function of time. $f_{\rm 300}$, for example, is the volumetric fraction of completely burned fuel within a radius of $300~{\rm km}$.  
        Left: comparing simulations without a magnetic field (wob1a) and a large-scale dipole field (dip3a). Right: comparing simulations with turbulence and turbulent magnetic fields that have medium-scale turbulence (turb1) with smaller-scale turbulence (turb4b) and also a strong magnetic field (turb4c). See Table~\ref{tab:prop} for simulation details. Small-scale preexisting turbulence is required for a more complete burning ($f$ values closer to 1), while a sufficiently strong magnetic field additionally increases the efficiency of burning and the burning rate.}
    \label{fig:fill_t}
\end{figure*}

We show in Figure~\ref{fig:fill_t} the filling factors within three radii of $300~{\rm km}$, $500~{\rm km}$, and $800~{\rm km}$, $f_{\rm 300}$, $f_{\rm 500}$, and $f_{\rm 800}$, respectively, as a function of time for selected simulations. Without turbulence, the filling factors are at most $0.2$ in the inner regions, and even smaller in the outer regions (left panel). A magnetic dipole field inhibits the propagation along its direction, and therefore results in even a lower filling factor. We also plot the maximal distance of burned material (the location of the deflagration front) along the direction of the dipole magnetic field, $z_{\rm max}$, and in the perpendicular direction, $R_{\rm max}$, as a function of time, displaying the squeezing effect again.

Turbulence fields, on the other hand, increase mixing and thus increase the filling factors. However, to efficiently mix the burned and unburned matter, the turbulence must be on a scale smaller than the pockets. Therefore, simulation with $r_{\rm D}=80~{\rm km}$ as in turb1, still develops plumes of burned matter with pockets between them, and the filling factors are higher but not close to 1. In turb4b, small-scale turbulence ($r_{\rm D}=40~{\rm km}$) efficiently drags the burned fuel into the pockets, allowing fuel consumption in the pockets. At about 0.7 seconds, the burning within 800 km is nearly complete. 

In addition, the strong turbulent $B$ field of several percent of the equi-partition in the simulation turb4c further mixes the unburned matter due to the Lorentz force. This results in very efficient burning, and complete burning is reached earlier in this simulation.

\begin{figure*}
\centering  
    \includegraphics[scale=0.45]{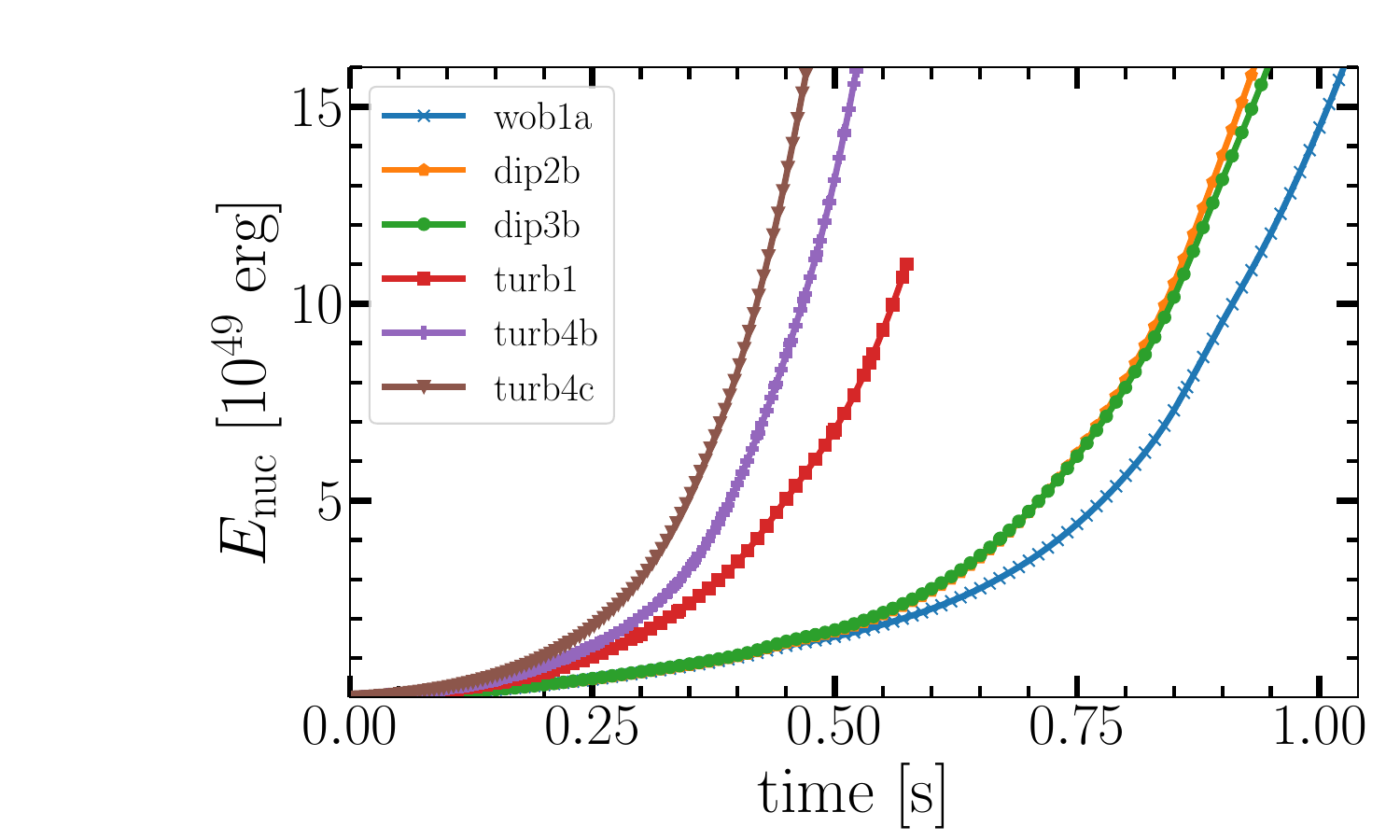}
    \includegraphics[scale=0.45]{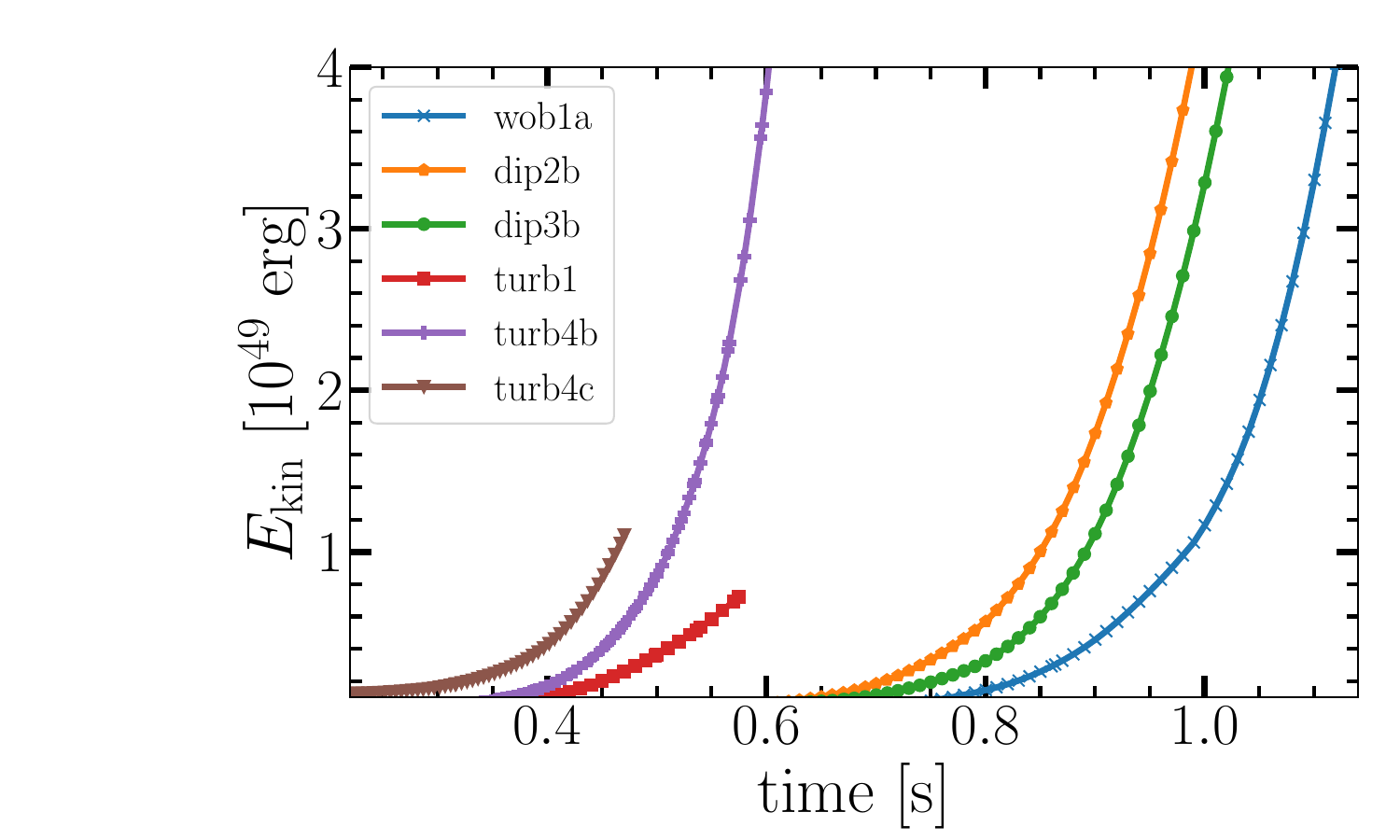}
    \caption{ Top: nuclear energy generation as a function of time. Bottom: Kinetic energy. Note that simulations 2 and dip2b have very little difference. A faster generation of nuclear energy leads to a faster WD expansion, causing the EC elements production to shut off at an earlier time.}
    \label{fig:energies}
\end{figure*}

Figure~\ref{fig:energies} shows the total nuclear energy generated (upper panel) and the total kinetic energy (lower panel) as a function of time in our simulations.
The small-scale turbulence in simulation turb4b increases the burning rate, and the total generated nuclear energy rises to $1.5\times10^{50}~{\rm erg}$, a third of the WD binding energy, slightly after $0.5~{\rm s}$. As the generated nuclear energy becomes comparable to the binding energy, the inner parts of the WD are lifted and start to expand, increasing the total kinetic energy (lower panel). The strong turbulent magnetic field in turb4c further increases the burning rate, and an energy of $1.5\times10^{50}~{\rm erg}$ is achieved before $0.5~{\rm s}$.

Simulations with larger turbulence sizes, turb1 and dip3b, are less efficient in filling the pockets and, therefore, result in lower burning rates. Simulation wob1a without any turbulence has the lowest burning rate, and an energy of $1.5\times10^{50}~{\rm erg}$ is reached only after $1~{\rm s}$. Interestingly, the strong dipole field of simulation dip3b restrains the WD expansion compared to dip2b with a 1000 times smaller $B$ field. 

As we show in Section~\ref{subsec:ec}, the expansion of the WD significantly alters the production of EC elements. This is because, as the WD expands, the densities decrease and EC reactions shut off. The earlier the expansion, the shorter the duration of active EC production.

\subsection{\label{subsec:ec} Influence on the production of Electron-Capture Elements}

Here, we employ the HYDRA module  for nuclear reactions   
(\citealt{Hoeflich2009} and references therein). As a reference model
\cite{Hoeflichetal2017}, we use
spherical, co-moving frame simulations with the same density and structure WD as the 3D simulations presented in this work. 

Nuclear reactions are calculated in the high-density regime at
temperatures above $5 \times 10^5\,{\rm K}$. We utilized a reaction network
whose rates include weak, strong, and electromagnetic reactions. The
network is based on the implementation of \citet{T94a,T94b}, but adopts
modified matrix solvers. For an isotope $i$, the change in the abundance
per nucleon, $Y_i$, is
\begin{equation}
    \dot Y_i =
    \sum_j N^i_j \lambda_j Y_j
    + \sum_{j,k} \frac{N^i_{j,k}}{1+\delta_{jk}}
    \rho N_A \langle \sigma v \rangle_{j;k} Y_j Y_k
    + O(3).
\end{equation}

The first term on the right-hand side includes single-particle
processes: decays, photo-disintegrations, electron and positron captures,
and neutrino-induced reactions, where $\lambda_j$ is the rate per
particle $j$. The second term includes two-particle reactions involving
particles with densities $n_j$ and $n_k$ in a plasma of mass density
$\rho$, where $N_A$ is Avogadro's number and $\delta_{jk}$ is the
Kronecker delta. The quantity $\langle \sigma v \rangle_{j;k}$ is the
nuclear cross section convolved with the relative velocity distribution
of particles $j$ and $k$ in the plasma. The coefficients $N^i_j$ and
$N^i_{j,k}$ denote the numbers of particles of species $i$ created or
destroyed in the corresponding reactions. The term $O(3)$ represents
contributions from multi-particle reactions.

Here we use 218 isotopes ranging from H to Zn. For temperatures above $6.5 \times 10^9\,{\rm K}$, we
assume nuclear statistical equilibrium, mediated by strong and
photon-induced reactions; only weak-interaction terms remain in the rate
equations. For $\langle \sigma v \rangle_{j;k}$, we assume a Boltzmann
velocity distribution for the isotopes in the plasma. Individual
reaction rates are fitted as functions of $\rho$ and $T$, with
coefficients tabulated in the REACLIB reaction-network library, together
with modified weak-reaction rates. Updated cross sections are taken from
\citet{Cy10}. For more details, see, e.g., \citet{T94a,Langanke04}.

As reference, we use a spherical delayed detonation model in the comoving frame throughout the explosion, up to the phase of free, homologous expansion. The initial WD model is the same as in the 3D series for the early deflagration phase \cite{Hoeflichetal2017}.
Its deflagration speed in mass coordinates is given by
\begin{equation}
v_t=max(v_{\rm cond},C_1 \sqrt{\alpha_T g L_f}) 
\end{equation}
with the Atwood number $\alpha_T = (\alpha -1)/(\alpha +1)$, where $\alpha =\rho^+
/ \rho^-$, $v_t \approx v_{\rm exp}=L_f~t$, and $v_{\rm exp}$ the expansion velocity at time $t$. 
$C_1$ has been calibrated \citep{dominguez00} based on 3D hydrodynamical simulations without pre-existing turbulence \citep{khokhlov97,Gamezoetal2003}. The radial resolution is 910 and $\approx 100$ points for the spherical and 3D models, respectively. 
The evolution of the burning rate as a function of time is very close to model wob1 within $\approx 2...3\% $, consistent with the shorter time and the lower resolution. 

The final abundances (masses) of EC elements (in $M_{\rm Ch}$) are given in Table~\ref{tab:ec2}. For the central density of $\rho_c=10^9~{\rm g~cm^{-3}}$, the main EC products are stable $^{54}{\rm Fe}$, $^{57}$Co/Fe and $^{58}$Ni. The $^{58}$Ni yields are lower for smaller $\rho_c$, while 
for increasing $\rho_c$, additional isotopes are produced, namely $^{48,42}$Ca, $^{52-55}$Cr, $^{55}$Mn, $^{52-60}$Fe, $^{58-62}$Ni, and Zn. \cite{Brachwitzetal2000,Hoeflich2018,2019A&A...630A..76G}.
Considering $\rho_c=10^9~{\rm g~cm^{-3}}$ allows us to avoid this shift in production from the $^{54}{\rm Fe},~^{57}{\rm Co},~^{58}{\rm Ni}$ group to lower isotopes when the baryon fractions drop to much lower values of $Y_e \leq 0.48$, and therefore better distills the influence of 3D effects on nuclear synthesis. Such a shift will otherwise introduce non-monotonicity in the individual EC production.  {At the densities considered, therefore, we identify the $^{58}$Ni mass relative to the reference model $f(^{58}{\rm Ni})\equiv M_{\rm 3D}^{^{58}{\rm Ni}} / M_{\rm ref,1D}^{^{58}{\rm Ni}}$ as representative of the EC production.}
\begin{table*}
\centering
\begin{tabular}{l|cccccc}
\hline
Isotope & $^{52}{\rm Cr}$ & $^{53}{\rm Cr}$ & $^{54}{\rm Cr}$ & $^{55}{\rm Mn}$ & \textbf{$^{54}$Fe} & \textbf{$^{57}$Fe}  \\
\hline
Mass [$M_{\rm ch}$] & 2.236 $(10^{-4})$ & 5.418 $(10^{-6})$ & 1.255 $(10^{-6})$ & 6.145 $(10^{-4})$ & \textbf{ 1.123 $(10^{-2})$} & \textbf{ 7.084 $(10^{-3})$ } \\
\hline
Isotope & \textbf{$^{59}$Co} & $^{52}{\rm Ni}$ & \textbf{$^{58}$Ni} & $^{60}{\rm Ni}$ & $^{61}{\rm Ni}$ & $^{62}{\rm Ni}$ \\
\hline
Mass [$M_{\rm ch}$] & \textbf{1.001 $(10^{-3})$} & 2.350 $(10^{-7})$ & \textbf{ 6.221 $(10^{-2})$} & 7.345 $(10^{-4})$ & 3.098 $(10^{-7})$ & 3.218 $(10^{-6})$ \\
\end{tabular}
\caption{Production of (stable) Electron Capture elements in the spherical 
reference model from \citealt{Hoeflichetal2017} of a near $M_{\rm Ch}$ mass WD with a central density of $10^9~{\rm  g~cm^{-3}}$. The initial WD has solar metallicity and originated from a star with a main-sequence mass of 7 $M_\odot$. The main isotopes (in bold face) are  $^{54/57}{\rm Fe}$, $^{59}{\rm Co}$, and $^{58}{\rm Ni}$. For larger $\rho_c$, the EC abundances shift toward 
more neutron reach elements, e.g. {Mn, and Cr} \cite{2019nuco.conf..187H,Brachwitzetal2000}.}
\label{tab:ec2}
\end{table*}

The main 3D effects are related to a change in the duration in which EC occurs, or better, to the decrease of the central density as a result of nuclear energy production and the acquisition of kinetic energy/expansion of the WD (Figure~\ref{fig:energies}). Without magnetic fields and turbulence (as in simulation wob1a), the duration of the electron capture phase is about 1.1 seconds with $f(^{58}{\rm Ni})=0.98$. 

Models with strong small-scale turbulence, consistent with turbulence created during a smoldering phase,
and $B_{\rm rms} \leq 3\times10^{11}~{\rm G}$  give $f(^{58}{\rm Ni}) \approx 0.46 $ (simulations turb4a and turb4b). 
The yields are hardly affected as long as B is well below
the saturation pressure. Stronger $B$ fields, in which the field reaches approximately 7\% of the saturation pressure ($B_{\rm rms}  \approx 5 \times 10^{12}$G), lead to significant mixing
of burned and unburned material in the central region (Figure~\ref{fig:fill_t}). As a result, $f(^{58}{\rm Ni})$ drops to $ \approx 0.42$ in simulation turb4c. 
Similarly, large-scale turbulence only drags the material rather than mixes it (wob2), leading to higher $f(^{58}\rm{Ni})$
compared to simulations of small-scale turbulence (turb4a).

Initial dipole B-fields and low turbulent velocities, $B\approx 10^{9,12}~{\rm G}$, lead to a lower reduction in $f(^{58}{\rm Ni})$ compared to small-scale strong turbulence, namely simulations dip2b and dip3b, respectively. The initial dipole B-field mostly drags the EC region rather than causing a more effective mixing of burned and unburned matter close to the central region (Fig.~\ref{fig:fill_t}). As a consequence, strong dipole $B$ fields (dip3b) have the opposite effect compared to turbulent fields; they restrain the effect of the turbulence, delay the expansion, and yield higher $f(^{58}{\rm Ni})$ compared to weaker dipole fields (dip2b). We note, however, that the resolution of these dipole models is relatively low. 

The main results are: in the most realistic scenario, 3D effects change the production of EC elements by about a factor of 2. As a consequence, the values of the central densities found by spectral analyzes based on spherical near-$M_{\rm Ch}$ models need to be increased by about $40\%$ \cite{hoeflich2006b,Diamondetal2015,Telescoetal2015,Diamond2018,2019A&A...630A..76G,DerKacyetal2024,Ashalletal2024,Kumaretal2026}.  {Specifically, the $\rho_c$ series of spherical models were used to find the $\rho_c $ with the same $^{58}{\rm Ni}$ as the 3D model, and their ratio
was used to estimate the increase in $\rho_c $ required, although it is also consistent with the EC yields proportional to the density square (assuming everything else remains constant).} 

Within the near-$M_{\rm Ch}$ class of scenarios, the range in $\rho_c$ is a factor of $\approx 5$. However, a factor of two may provide an alternative interpretation of the finding that $^{58}$Ni mass increases
with decreasing brightness and instead points to a systematic change in the progenitor system. We also have to note that high-density burning can be achieved
within the scenario of secular mergers and core-degenerate scenarios \cite{hk96,KashiSoker2011,Hoeflich2017book,Lu:2023}.
In particular, at higher densities, additional low-$Y_e$ isotopes are expected to be detected in SNe Ia with JWST, posing a challenge to nuclear data. The current work should be 
regarded as a first step only and requires more extensive studies in the future.

\section{\label{sec:conclusions} Conclusions}

Thermonuclear supernovae play a major role in modern cosmology. Advances in simulations and observations allow unique tests of detailed physics by high-precision spectroscopy to a level of $\approx 10\% $ (see, e.g., the mid-IR spectra shown in \citealt{Ashalletal2024}).
This increase in accuracy requires an adequate improvement in the precision of nuclear physics, namely EC rates on IGE and on Mn and Cr, well beyond the accuracy that was sufficient only a few years ago.

Based on 3D MHD simulations, we present a detailed study of the early deflagration phase in near-$M_{\rm Ch}$ models and the production of EC elements in this phase. We additionally compare the 3D results with spherical simulations commonly used in the literature. The following conclusions are obtained:

\begin{itemize}
    
     \item 
       Promising near-$M_{\rm Ch}$ models based on spherical simulations for the explosions allow us to reproduce all observables. 
       However, starting from a static WD, full 3D models result in large unburned pockets, and the released nuclear energy is deposited mainly in a few burned rising plumes in an otherwise globally static WD. In contrast to 1D models, the unburned pockets of C/O reside throughout the entire WD, and no substantial pre-expansion of the WD occurs during the deflagration. This results in a negligible amount of IME produced, which is in disagreement with observations \cite{Nomoto1984}. Note that pure deflagration models have been found deficient in reproducing pre-maximum spectra, while models with a transition to detonation \cite{khok89} are consistent with modern observations. 
       
    \item 
       We demonstrated that preexisting small-scale turbulent fields, as expected from thermonuclear burning prior to the explosion, burn away the pockets, effectively restoring the properties found in spherical simulations (Section~\ref{subsec:mhd}).

    \item  
        We show that $B$ fields become important during the early deflagration phase if B exceeds $\approx 1\%$ of the saturation field. We note, though, that evidence for these ultra-high B fields is growing (Section~\ref{sec:intro}).

    \item 
        The production of EC elements mostly depends on the central density of the WD. For observed SNe Ia, the central densities within the framework of spherical models range from $\approx 8 - 50 \times 10^8~{\rm g~cm^{-3}}$, producing between 0.01 to 0.15 $M_\odot$ of $^{58}$Ni, respectively   \cite{Brachwitzetal2000,2019nuco.conf..187H,Hoeflichetal2017}.

    \item 
       We find that 3D effects on the EC require a systematic increase of $\approx 40\% $ in $\rho_c $ relative to the results based on spherical models. Turbulence and turbulent magnetic fields regulate whether the pockets are burned, the starting point of the WD expansion, and consequently, the duration of active EC production. $Y_e$ depends on the nuclear rate and the duration of the high-density burning phase (Figure~\ref{fig:energies}).

     \item 
        Dipole fields may be produced during a prolonged accretion phase of merging WDs. We showed that dipole B fields redistribute central burning products but are unable to efficiently burn the pockets.

\end{itemize}

Finally, we want to emphasize the limitations of the current study. First, our simulations are limited to the 
early deflagration phase. Though this is sufficient to determine the EC masses, the final
distribution in velocity space may be affected by the subsequent RT dominated deflagration and detonation phase. Future simulations that include the subsequent deflagration and transition to detonation will answer whether the large filling factor in the inner EC region produces a barrier for RT going inward.
We also intend to extend the simulations to incorporate a broader regime of central densities, from commonly found WDs with low central densities \cite{2004ApJ...617.1258H,Telescoetal2015,Diamondetal2015,Ashalletal2024,Kumaretal2026} to ultra-high densities
close to the Accretion Induced Collapse limit for neutron star formation, as evidenced by observations \cite{2019A&A...630A..76G}.
 {In addition, we further expand the grid of parameters that describe our initial conditions in a forthcoming paper. We note that in some of the simulations, the initial magnetic field experiences growth during the simulated evolution. However, a detailed study of this effect is left for a future paper.}
Furthermore, secular mergers will be studied in the future as another alternative for scenarios with high-density burning.

\begin{acknowledgments}
 {We thank the referees for their useful comments that helped improve the quality of the paper.} Sa.S. and P.H. acknowledge the support from the National Science Foundation NSF awards AST-1715133 and  AST-2306395 for enabling the development of the methods and computational codes. P.H., C.A., and J.M.D acknowledge support from NASA grants JWST-GO-02114,
JWST-GO-02122, JWST-GO-04522, JWST-GO-04217, JWST-GO-04436,
JWST-GO-03726, JWST-GO-05057, JWST-GO-05290, JWST-GO-06023,
JWST-GO-06677, JWST-GO-06213, JWST-GO-06583, JWST-GO-09231.
The software used in this work was developed in part by the DOE NNSA- and DOE Office of Science-supported Flash Center for Computational Science at the University of Chicago and the University of
Rochester. The line profiles were calculated with the radiation-hydro code HYDRA.
This work also required using and integrating a Python package for astronomy, yt (https://yt-project.org, \citealt{Turk2011}).
\end{acknowledgments}

\section*{Data Availability Statement}

The data that support the findings of this study are available from the corresponding authors upon reasonable request.

\appendix

\section{\label{appendix:mhd-eqs} MHD equations}
We use \texttt{FLASH}\cite{Fryxelletal2000} to solve the Eulerian equations of ideal MHD. In c.g.s units they have the form:

\begin{equation} 
\label{eq:MHD1}
\mypd{\rho}{t}
+ \nabla \cdot
(\rho \mathbf{v})
= 0,
\end{equation}

\begin{equation} \label{eq:MHD2}
\mypd{\rho \mathbf{v} }{ t}
+ \nabla \cdot
\left( 
	\rho \mathbf{v} \mathbf{v} 
	+ \mathbf{I} P 
	- \frac { \mathbf{B} \mathbf{B} }{ 8 \pi } 
\right)
=
 \rho \mathbf{g},
\end{equation}

\begin{equation} \label{eq:MHD3}
\mypd{\rho E}{t}
+ \nabla \cdot
\left[
	(\rho E + P) \mathbf{v} 
	-
	\frac 
	{ \mathbf{B} ( \mathbf{B} \cdot \mathbf{v} ) }
	{ 4 \pi }	
\right]
=
 \rho \mathbf{v} \cdot \mathbf{g} 
+ \rho \dot{E}_{\rm nuc},
\end{equation}

\begin{equation} \label{eq:MHD4}
\mypd { \mathbf{B} }{ t}
+ \nabla \times ( \mathbf{v} \times \mathbf{B})
=0,
\end{equation}
where, $\mathbf{v} \mathbf{v}$ and $\mathbf{B} \mathbf{B}$ are 
the velocity and the magnetic field outer products,
$\rho$, $\mygvec$, and $\dot{E}_{\rm nuc}$ are the density, 
the gravitational acceleration, 
and the rate of specific energy production from the nuclear burning.
Further 
$E = e + {v^2 } / { 2 } + {B^2} / {8\pi}$
is the sum of the specific internal energy of the gas ($e$), the specific kinetic energy of the gas, and magnetic energy, and 
$P = p + {B^2} / {8\pi}$
is the sum of the gas pressure $p$ and magnetic pressure.

The Helmholtz equation of state \cite{TimmesSwesty2000} (EOS) is used to close the system of equations, where the gas specific internal energy and pressure are calculated as the sum over the components
\begin{equation} \label{eq:EOS1}
e = e_{\rm rad} +e_{\rm ion} + e_{\rm ele} + e_{\rm pos} + e_{\rm coul},
\end{equation}
and
\begin{equation} \label{eq:EOS2}
p = p_{\rm rad} +p_{\rm ion} + p_{\rm ele} + p_{\rm pos} + p_{\rm coul}.
\end{equation}
Here the subscripts “rad,” “ion,” “ele,” “pos,” and “coul” represent the contributions from radiation, nuclei, electrons, positrons, and corrections for Coulomb effects, respectively. The radiation portion assumes a blackbody in local thermodynamic equilibrium, the ion portion (nuclei) is treated as an ideal gas with $\gamma = 5/3$, and the electrons and positrons are treated as a non-interacting Fermi gas (see \citealt{Fryxelletal2000} for more details on how this is actually calculated in \texttt{FLASH}). 

The gravitational acceleration, $\mathbf{g}$, is computed using \texttt{FLASH} Poisson multipole solver with a highest multipole order of 16 and isolated boundary conditions.

Because the physical width of the burning front
is small compared to our grid resolution, we approximate the burning propagation using a flame-capture technique \cite{Khokhlov1995}. The burned mass fraction $\phi$ therefore obeys a reaction-diffusion equation
\begin{equation} \label{eq:flame-diffusion}
\mypd{\phi}{t} + \mathbf{v} \cdot \nabla \phi
= \kappa \nabla^2 \phi + R\left(\phi \right)/\tau,
\end{equation}
where $\phi=0$ is pure fuel. $\kappa$ and $R\left(\phi \right)/\tau$ are the diffusion rate and reaction rate,
respectively, and $R$ adopts the sharpened Kolmogorov Petrovski Piskunov (sKPP) form:
\begin{equation} \label{eq:reaction-term}
R
= \frac{f_b}{4} \left(\phi - \varepsilon_0 \right)\left(1 - \phi + \varepsilon_1 \right).
\end{equation}

As in \citealt{Vladimirova2006}, we scale the diffusion and reaction terms $\kappa = v_{\rm burn}b\Delta x/16$ and $\tau = b\Delta x/\left(16 v_{\rm burn}\right)$ according to a prescribed speed $v_{\rm burn}$ and numerical front width of several cells, where $\Delta x$ is the size of a grid cell. 
The default values for the parameters $\varepsilon_0=~\varepsilon_1=10^{-3},~f_b=1.309$,~and~$b=3.2$ were chosen \cite{Townsley2007}.

We assume instantaneous burning within the flame front, releasing specific energy at a rate of $\dot{E}_{\rm nuc}=\dot{\phi}Q_{\rm nuc}$, where $Q_{\rm nuc}$ is the release of nuclear (specific) energy inside the front. 
We set the nuclear burning according to the local density. Nuclear statistical equilibrium (NSE) $Q_{\rm burn}=Q^{\rm NSE}=7.2\times10^{17}~{\rm erg~g^{-1}}$, nuclear statistical quasi-equilibrium (NSQE) $Q_{\rm burn}=Q^{\rm NSQE}=6.1\times10^{17}~{\rm erg~g^{-1}}$, and carbon burning $Q_{\rm burn}=Q^{\rm C}=2.5\times10^{17}~{\rm erg~g^{-1}}$ are assumed for densities $\rho>2\times 10^{7}{\rm ~g~cm^{-3}}$, $2\times 10^{7}{\rm ~g~cm^{-3}}\ge \rho > 10^{6}{\rm ~g~cm^{-3}}$, and  $10^{6}{\rm ~g~cm^{-3}}\ge \rho$, respectively. However, in this study, we focus  on the early phase of the deflagration, where the front remains in high-density regions, and the main burning is through NSE, while NSQE and carbon burning contribute only marginally. 

\bibliography{refs,article_add,article_mod,article1,articlesnew}

\end{document}